\documentclass[aps,pra,reprint,superscriptaddress]{revtex4-2}

\usepackage[T1]{fontenc}
\usepackage{graphicx}
\usepackage{verbatim}
\usepackage[separate-uncertainty=true]{siunitx}
\usepackage{amsmath}
\usepackage{physics}
\usepackage{color}

\usepackage{hyperref}
\hypersetup{%
	colorlinks,
	citecolor=blue,
	linkcolor=blue,
	urlcolor=blue
}

\newcommand{\Nth}{N_\mathrm{th} }
\newcommand{\Rb}{$^{87}$Rb}
\newcommand{\figref}[1]{Fig.~\ref{#1}}

\makeindex
\begin{document}

\title{Atom Number Fluctuations in Bose Gases -\\ Statistical analysis of parameter estimation}

\author{T.\,Vibel}
\author{M.\,B.\,Christensen}
\author{R.\,M.\,F.\,Andersen}
\author{L.\,N.\,Stokholm}
\affiliation{Center for Complex Quantum Systems, Department of Physics and Astronomy, Aarhus University, Ny Munkegade 120, DK-8000 Aarhus C, Denmark.}
\author{K. Paw\l{}owski}
\author{K. Rz\k{a}\.zewski}	
\affiliation{Center for Theoretical Physics, Polish Academy of Sciences, Al. Lotnik\'{o}w 32/46, 02-668 Warsaw, Poland.}
\author{M.\,A.\,Kristensen}
\author{J.\,J.\,Arlt}
\affiliation{Center for Complex Quantum Systems, Department of Physics and Astronomy, Aarhus University, Ny Munkegade 120, DK-8000 Aarhus C, Denmark.}

\begin{abstract}
The investigation of the fluctuations in interacting quantum systems at finite temperatures showcases the ongoing challenges in understanding complex quantum systems. Recently, atom number fluctuations in weakly interacting Bose-Einstein condensates were observed, motivating an investigation of the thermal component of partially condensed Bose gases. Here, we present a combined analysis of both components, revealing the presence of fluctuations in the thermal component. 
This analysis includes a comprehensive statistical evaluation of uncertainties in the preparation and parameter estimation of partially condensed Bose gases. Using Monte Carlo simulations of optical density profiles, we estimate the noise contributions to the atom number and temperature estimation of the condensed and thermal cloud, which is generally applicable in the field of ultracold atoms. 
Furthermore, we investigate the specific noise contributions in the analysis of atom number fluctuations and show that preparation noise in the total atom number leads to an important technical noise contribution. Subtracting all known noise contributions from the variance of the atom number in the BEC and thermal component allows us to improve the estimate of the fundamental peak fluctuations.
\end{abstract}

\maketitle

\tableofcontents
%###############################################################################
% ------------------ Introduction ------------------------------ 
%###############################################################################
\section{Introduction}
Bose-Einstein Condensation (BEC) in dilute gasses has become a fundamental tool in the exploration of quantum physics and presents notable potential for advancements in several quantum technologies such as quantum simulators~\cite{Bloch2012} and quantum sensors in metrological applications~\cite{Bongs2019}. Therefore, a comprehensive understanding of the statistical errors involved in the preparation and detection of BECs is crucial to ensure the validity of the results based on it. This understanding is particularly important for the analysis of fundamental atom number fluctuation in quantum degenerate Bose gases, which is the primary motivation of this work. 

The problem of the fundamental atom number fluctuations in ultracold Bose gases is a complex issue that led to a long-standing theoretical debate started by E. Schr\"odinger and continued by the next generation of physicists \cite{Ziff1977, Politzer1996, Kocharovsky2006}. A central issue in these discussions was the question which statistical ensemble should be employed in the theoretical calculations. In particular, for an ideal gas trapped in a spherical trap, the variance of the atom number in the BEC calculated in the microcanonical ensemble is less than half of the result in the canonical ensemble even in the limit of infinitely many atoms~\cite{Navez1997, Holthaus1998}. Moreover, the commonly used grand canonical ensemble description of the non-interacting Bose gas leads to absurdly large fluctuations. This finding contradicts the typical notion that results should be independent of the chosen ensemble in the thermodynamic limit.

Therefore, one has to invoke the statistical ensemble which represents the experimental reality best to correctly describe the physical system. Bose-Einstein condensates in dilute atomic gasses are nearly perfectly isolated, weakly interacting systems and one naturally expects a microcanonical ensemble description to be most appropriate. This was indeed confirmed by recent experiments, which observed significantly lower variance of the atom number fluctuations than expected in a canonical result~\cite{Christensen2021}.

However, a quantitative, microcanonical calculation of the atom number fluctuations in an ultracold Bose gas with realistic atom numbers and interaction strengths is not available. Therefore, current theoretical investigations focus on the role of interaction and geometry. To guide these investigations it is important to achieve the most accurate experimental extraction of the atom number fluctuations, which motivates a detailed exploration of technical noise sources. 

Experimentally, the atom number fluctuations in partially condensed Bose gasses manifest themselves as variations of the atom number in the condensed and the thermal part of the atomic cloud under identical experimental conditions. Previous measurements detected these fluctuations~\cite{Kristensen2019} and characterized their dependence on atom number and trap geometry~\cite{Christensen2021}. For a fixed total number of atoms, the number of the condensed atoms is thus expected to be perfectly anticorrelated with the number of thermal atoms in Bose gas. Simply put, one expects atoms leaving the BEC must enter the thermal component and vice versa. This characteristic property of the fluctuations in a partially condensed Bose gas motivated our analysis of the thermal atom number variance and, consequently, the complete noise analysis of the system.

The noise analysis of the system is based on a detailed discussion of the imaging errors for ultracold atomic clouds and the resulting statistical errors when the experimental parameters of these clouds are extracted.

Prior investigations of ultracold quantum gases based on absorption imaging have focused on systematic errors and strategies to mitigate them. These include the calibration of the effective saturation intensity~\cite{Reinaudi2007,Hueck2017,Veyron2022,Vibel2024alpha}, the appropriate choice of fitting region for the thermal cloud~\cite{Szczepkowski2009}, and the effect of increased cloud width due to recoil heating and defocusing~\cite{Ketterle1999,Muessel2013}. Correction techniques for the Doppler shift during imaging have also been explored~\cite{Genkina2015,Hueck2017}. Experimental techniques have been developed to precisely determine the optimal focus for the imaging setup~\cite{Putra2014}. 

The statistical errors that occur in the parameter estimation of ultracold atomic clouds have only received relatively little attention. In experiments aiming at precise atom number determination, the optimal settings for imaging power, pulse duration, and cloud size to minimize the impact of photon shot noise~\cite{Pappa2011,Horikoshi2017,Genkina2015,Lueckes_PhD} were explored. Techniques in image processing, aiming to reduce the effects of photon shot noise and eliminate fringes due to vibrations, have also been investigated~\cite{Ockeloen2010,Niu2018, Ness2020, Song2020}. Statistical effects might not have received much attention in the past, since preparation errors of ultracold atomic clouds typically exceeded parameter estimation errors. However, current experiments have reached a regime where parameter estimation errors are relevant, as discussed in the cloud parameter noise analysis of the system.

The measurement of fundamental atom number fluctuations is particularly challenging in this respect since technical noise arising from the preparation and parameter estimation of ultracold atomic clouds  directly competes with the main experimental results~\cite{Kristensen2019,Christensen2021}. Here, we develop a detailed model of the technical noise sources which allows us to subtract their contribution from the observed variances and to isolate the fundamental atom number fluctuations. The Monte Carlo simulation employed to identify the specific noise source of estimation errors can be applied to all analyses of density profiles obtained from absorption imaging. 

The paper is organized as follows. Section~\ref{sec:ExperimentalSetup} presents the experimental setup and related image analysis techniques. The statistical noise analysis of the estimation errors for bimodal cloud parameters is presented in Sec.~\ref{Sec:CloudAnalysis}. The specific noise sources relevant for the measurement of atom number fluctuations are discussed in Sec.~\ref{sec:FlucsMeasurementNoise}. All technical errors are combined in Sec.~\ref{sec:FlucsMeasurementNoise} and allow for the evaluation of the fundamental atom number fluctuations. In Sec.~\ref{sec:conclusion} we conclude and provide an outlook on future experimental and theoretical developments in Sec.~\ref{sec:outlook}.

%###############################################################################
% ---------- Experimental description + data analysis -----------------
%###############################################################################

\section{Experimental description and data analysis}
\label{sec:ExperimentalSetup}

\subsection{Experimental setup}

The experimental apparatus was previously described in~\cite{Kristensen2019,Christensen2021}. Briefly, $\approx\!10^9$~\Rb{} atoms are collected and cooled with a magneto-optical trap. The atoms are then optically pumped to the $\ket{F = 2, m_F = 2}$ state and transported to another part of the vacuum chamber with a lower pressure. Here, the atoms are transferred to a harmonic magnetic trap with initial radial and axial trapping frequencies, $\omega_\mathrm{r}/2\pi = \SI{300}{\hertz}$ and $\omega_\mathrm{z}/2\pi  =\SI{17}{\hertz}$. Subsequently, the atoms are cooled in this trap configuration using forced radio-frequency (rf) evaporative cooling. Control of a current bypassing the quadrupole coils allows us to lower (primarily) the radial trapping frequency, and, thereby, access aspect ratios, $\lambda = \omega_\mathrm{r} / \omega_\mathrm{z}$, in a range from 17 to 4.5.

Before reaching the phase transition to a BEC, the atom number is stabilized using a combination of non-destructive imaging and loss pulses of varying duration. When the atom clouds contain $\sim 4\times 10^6$~atoms at a temperature \SI{14}{\micro \kelvin}, they are probed in-trap with off-resonant Faraday imaging~\cite{Kristensen2017}. Based on these images, the atom number is evaluated on a real-time operated field-programmable gate array (FPGA) and compared with a reference value. The FPGA calculates the duration of a weak rf loss pulse to remove any excess atoms. The result of the stabilization is verified in a second Faraday measurement. This technique allows us to control the atom number with a relative stability of $\approx\!10^{-4}$~\cite{Kristensen2017,Gajdacz2016}.

After stabilizing the atom number, the trap is smoothly decompressed to the final aspect ratio, $\lambda = 5.4$, and the cooling to reach BECs is completed. This value was picked as a compromise featuring appropriately high radial trapping frequencies while avoiding the presence of phase-fluctuations~\cite{Christensen2021,Shah2023}.  Equilibrium is ensured by holding the final radio frequency for 800~ms, after which the atoms are held for a further 400~ms without rf radiation before trap turn-off.  

To probe the system, the trap is abruptly turned off while maintaining a bias field, and the cloud expands in free fall for 35~ms. After this, it is imaged using resonant light with a square pulse of duration $\tau = \SI{30}{\micro \second}$ and an intensity of $I_0=3.2~I_\mathrm{sat}$, where $I_\mathrm{sat}$ is the saturation intensity. Imaging is applied perpendicular to the axis of the cigar-shaped trap and consists of two pairs of convex lenses and a camera. The first pair creates an intermediate image plane, which is imaged onto a CCD camera by the second lens pair. Four razor blades trim the beam in the intermediate plane such that the beam light only hits the camera chip. This hinders scattered beam photons from polluting the image and minimizes external stray light. The imaging light is circular polarized and resonant with the closed $\ket{F=2, m_\mathrm{F}=2}$  to $\ket{F'=3, m_\mathrm{F'}=3}$ D$_2$ transition. During the trap turn-off, the magnetic field is rotated from orthogonal to parallel with the imaging direction. 

The sequence consists of 6 images in total. First, an absorption image with atoms (probe) is acquired. Then, the atoms are optically pumped to the $\ket{F=1, m_\mathrm{F}=1}$ state to render them transparent for subsequent images. This allows an image without atoms (reference) to be taken only \SI{340}{\micro \second} later, limited only by the camera shift speed. As a result, the processed images of optical densities show no visible fringes due to vibrations. Two additional images are taken a few seconds later to account for dark counts and background light. Finally, two more images without atoms are captured, which are identical to the probe and reference images. These images are used for diagnostic purposes such as the analysis of pixel noise.

% ---------- Parameter extraction -----------------

\subsection{Determining atom cloud parameters}
\label{sec:cloudParams}

Based on the images of the atom clouds, the relevant parameters, such as atom number and temperature, are determined as follows. The optical density ($od$) and the column density $\tilde{n} = \int\mathrm{d}z\, n(z)$ for the camera pixel labeled $\ell$ is calculated using
\begin{align}
\label{eq:od}
od^{(\ell)} &= \tilde{n}^{(\ell)}\sigma_0 \nonumber \\ 
            &=  \alpha^*(1+4\tilde{\delta}^2)\ln\left(\frac{\tilde{I}^{(\ell)}_0}{\tilde{I}^{(\ell)}_\mathrm{p}}\right) + 
\tilde{I}^{(\ell)}_0-\tilde{I}^{(\ell)}_\mathrm{p}, 
\end{align}
where $\sigma_0$ is the resonant scattering cross-section and $\alpha^*$ is an imaging calibration factor~\cite{Reinaudi2007, Vibel2024alpha}. The detuning $\tilde{\delta}$ is typically set to zero in resonant absorption imaging, however, its inclusion in Eq.\ref{eq:od} is important for the evaluation of estimation errors. The probe and reference intensities are $\tilde{I}_\mathrm{p}$ and $\tilde{I}_\mathrm{0}$ respectively, in units of the saturation intensity $I_\mathrm{sat}$ These intensities are obtained from camera count values according to 
\begin{align} 
\label{eq:Intensity}
    \tilde{I}_j^{(\ell)} = \frac{I_j^{(\ell)}}{I_\mathrm{sat}} = \frac{\mathcal{N}_{\gamma,j}^{(\ell)} \hbar \omega}{I_\mathrm{sat} \tau A}	
    = \frac{\mathcal{N}_{\mathrm{el},j}^{(\ell)}}{\mathcal{T} \eta} \frac{\hbar \omega}{I_\mathrm{sat} \tau A}
    \notag\\
    = \frac{\mathcal{N}_{\mathrm{c},j}^{(\ell)}}{\mathcal{T} \eta g} \frac{\hbar \omega}{I_\mathrm{sat} \tau A}= C \cdot \frac{\mathcal{N}_{\mathrm{c},j}^{(\ell)}}{\tau}
\end{align}
where the index $j$ refers to the probe (p) or reference (0) image, $I$ is the real intensity, $\omega$ is the imaging frequency, $\tau$ is the imaging pulse duration, $A$ is the pixel area in the object plane, $\mathcal{T}$ is the transmission through the optical elements in the imaging system, $\eta$ is the quantum efficiency of the camera chip, and $g$ is the camera gain. The number of photons in the object plane, number of photo-electrons, and camera counts are denoted $\mathcal{N}_\gamma$,  $\mathcal{N}_\mathrm{el}$, and $\mathcal{N}_\mathrm{c}$, respectively. We assume that these images are dark-image corrected. The camera calibration factor is thus $C = \hbar \omega / (\mathcal{T} \eta g I_\mathrm{sat} A)$.

\begin{figure}
	\centering
	\includegraphics[width=0.99\columnwidth]{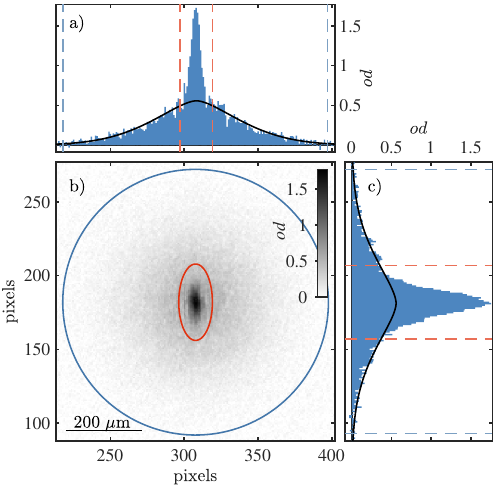}
	\caption{Example of an optical density profile (b) and cross sections through the centre in both directions, a) and c). The radius of the blue ellipse is 3 times the thermal width (as defined in Eq.~\ref{eq:Bose-enhanced}). The area within the blue ellipse is called the \textit{cloud region}. The area within the red ellipse is the \textit{BEC region}, and the area between the blue and red ellipse corresponds to the \textit{wings} of the thermal cloud. The edges of the cloud and BEC region are projected as dashed lines in the side panels. The black lines in a) and c) show the thermal fit.}
	\label{fig:cloudFitExample}
\end{figure}

Figure~\ref{fig:cloudFitExample} shows cross-sections of a typical optical density profile. The relevant parameters are extracted by fitting the wings of the cloud (area between the blue and red ellipse) with a Bose-enhanced Gaussian distribution. After integrating over the $z$-direction and in the low-temperature limit with the chemical potential set to 0, the distribution takes the form~\cite{Ketterle1999, Kristensen2018}
\begin{align}\label{eq:Bose-enhanced}
		\notag od_{\mathrm{BE}}(x, y) =& \frac{N_\mathrm{th} \sigma_0}{2 \pi w_x w_y \zeta(3)}\\
  &g_{2}\left(\exp\left(-\frac{(x -c_x)^2}{2w_x^2} -\frac{(y -c_y)^2}{2w_y^2}\right)\right),
\end{align} 
where the free fit parameters are the total number of thermal atoms $\Nth$, the widths $w_\mathrm{i}$, and the centre coordinates $c_\mathrm{i}$ for $i = \{x, y\}$. For our setup, this corresponds to $i = \{z, r\}$. The Bose function is $g_\gamma (z) = \sum_{n=1}^{\infty}\frac{z^n}{n^\gamma}$ and $\zeta(x)$ is the Riemann Zeta function. Given that the low-temperature limit of the Bose-enhanced function is the only analytic fitting function available in this regime, it is used across the entire temperature range. 

The widths are linked to the effective temperature, $T_i$, through the equation $w_\mathrm{i} = \sqrt{k_\mathrm{B}T_i/m } \sqrt{ 1/\omega_\mathrm{i}^2+t^2}$.  Here, $\omega_\mathrm{j}$ for $i = z, r$ denotes the axial and radial trapping frequencies, $t$ is the time of free expansion, and $m$ is the mass of the atomic species. These temperatures are averaged to a single temperature according to \cite{Gerbier2004,Szczepkowski2009}.

In a second step, the fit is extrapolated to the BEC region (red ellipse in Fig.~\ref{fig:cloudFitExample}) and subtracted from the data. The remaining signal within this region is then summed up to determine the BEC atom number, $N_0$. Additionally, the total atom number $N$ is measured by summing the signal in the entire cloud region according to $N = A/\sigma_0 \sum_{\ell\in \mathrm{R}} od^{(\ell)}$, within the respective region \emph{R}.

The final evaluation of all clouds in the analysis uses a fixed centre of the cloud region and a fixed elliptical BEC region large enough to encompass the largest BEC. The radii of the BEC region are set to $S=1.13$ times the largest recorded BEC radii from Thomas-Fermi fits (red ellipse in Fig.~\ref{fig:cloudFitExample}). Additionally, a slightly larger elliptical region defined by $S=1.2$ sets the inner boundary for the thermal fit region to ensure a stable fit~\cite{Szczepkowski2009}. 

The outer boundary of the wings (blue ellipse) is determined from a pre-analysis of the entire fluctuations data set performed with a fixed thermal region large enough to include the warmest cloud. The total atom number, $N$, and the widths, $w_j$, are extracted for all clouds. We fit a higher-order polynomial to $w_i$ as a function $N$, which provides us with $w_i(N)$. In the final evaluation, we set the cloud region to $3\ w_i(N)$ based on the total atom number $N$ in an image to be evaluated.

\subsection{Measurement of atom number fluctuations}
\label{sec:fluctmeasure}

\begin{center}
	\begin{figure*}
		\centering
		\includegraphics[width=\textwidth]{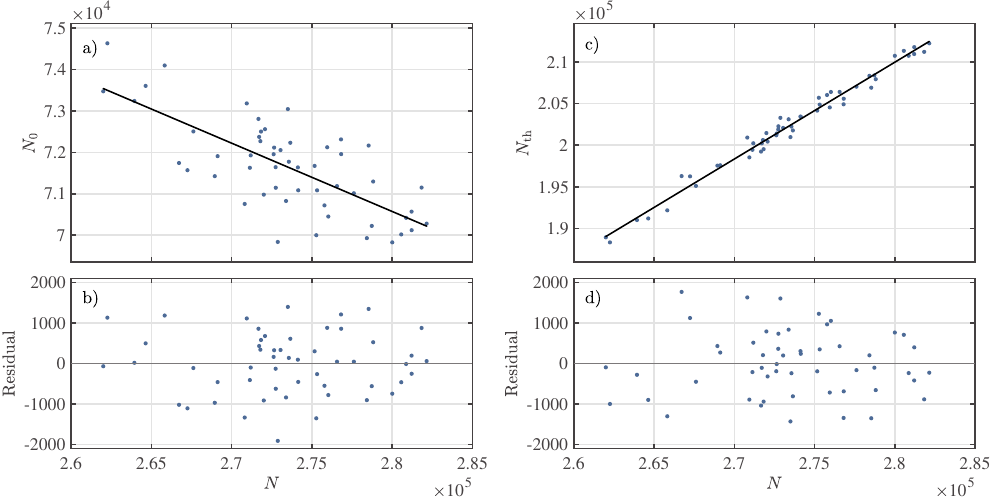}
		\caption{In order to prevent technical drift caused by the stochastic nature of evaporation, $N_0$ (a) and $\Nth$ (c) are plotted against $N$ and fitted to a line. The variances of these are calculated from the residuals shown in (b) and (d).}
		\label{fig:residual}
	\end{figure*}
\end{center}

The fluctuations are measured as a function of temperature (corresponding to a final rf). To determine fluctuations at a specific temperature, the variance of the atom number in the BEC and the thermal cloud needs to be obtained from samples prepared under the same conditions. In practice, $\approx\! 60$ experimental realizations were used to obtain a variance. Only realizations for which the atom number stabilization succeeded were included, and a minimum of 45 clouds were required to evaluate the variance. Figure~\ref{fig:residual} shows the atom number in the BEC, $N_0$, and the thermal atom number, $\Nth$ as a function of the total number $N$ for such a set of measurements.

Since each measurement takes $\approx\!90$~s, a measurement set takes considerable time during which the experimental apparatus is exposed to drifts in temperature. This leads to small drift of the trap depth, which results in a small drift in the total atom number and temperature. Consequently, the variance of the atom numbers cannot be calculated directly. Instead, the technical drift within a set of $\approx\! 60$ experimental realizations is eliminated by subtracting a linear fits $N_0(N)$ from the BEC atom number and $N_\mathrm{th}(N)$ from the thermal atom number as shown in \figref{fig:residual}. This subtraction proved vital to the observation of fluctuations~\cite{Kristensen2019,Christensen2021} and in the following, it is referred to as the \textit{correlation method}.

The variance is then determined as the two-sample variance~\cite{Hume2013} of the residuals shown in \figref{fig:residual}~(b) and (d) 
\begin{equation}\label{eq:TwoSampleVariance}
	\Delta x^2 = \frac{1}{2}\left\langle \left(x_{j+1} - x_j\right)^2 \right\rangle.
\end{equation}
Here, $x_{j+1}$ and $x_j$ represent two adjacent measurements (in time). The two-sample variance includes variance in atom number and detection noise but excludes slower technical drifts. The error of the variances is given by~\cite{Lueckes_PhD,cho2009} 
\begin{equation}\label{eq:varianceError}   
    e(\Delta x^2) = \sqrt{n\frac{m_4(n - 1)^2 - m_2^2(n^2 - 3)}{(n - 3)(n - 2)(n - 1)^2}}
\end{equation}
where $m_k$ are the central moments of the sample given by $m_k = 1/n \sum_i^n \left( x_j - \overline{x} \right)^k$ and $n$ is number of measurements.

Figure~\ref{fig:fluctuations} shows the measured variances of the BEC (red dots) and the thermal cloud (blue dots), which is in detail in Sec.~\ref{sec:technoiseparam}. In addition, the temperature variance was studied. Again, the two-sample variance of the residual was taken after subtracting a linear fit from the temperatures $T$ as a function of $N$ for each of the sets of $\approx\! 60$ clouds. Figure~\ref{fig:temperatureSpread} shows the standard deviation of $T$ as a function of temperature. It lies around \SI{1}{\nano \kelvin} and slightly increases toward colder temperatures, where smaller clouds of atoms degrade the thermal fit.

In both cases, it is clear that a detailed model of technical noise and its effects on the extracted variances is of great importance. Such a model of the inherent technical noise sources is a main result of this paper, as discussed in the following sections.

% ###############################################################################
% ------------------ Cloud parameter noise analysis ------------------------------
% ###############################################################################

\section{Cloud parameter noise analysis}
\label{Sec:CloudAnalysis}

This section focuses on the effect of technical noise on the estimation errors associated with cloud parameters. The mechanisms responsible for these errors are general and applicable to any analysis of optical density profiles obtained using absorption imaging, not just studies of fundamental atom number fluctuations. Due to the complexity of the data analysis, it is challenging to derive analytical expressions for the estimation errors, and the final analysis is based on a Monte Carlo simulation.

In the following, the most critical mechanisms impacting the estimation errors and their implementation in the Monte Carlo simulation are described. In addition, we briefly discuss the imaging effects not included in the simulation.

% ------------ Estimation errors ------------------
\subsection{Imaging noise in optical density}
\label{sec:odNoise}

The main contributions to uncertainties in the value of the optical density in each pixel of the absorption images (Fig.~\ref{fig:cloudFitExample}) arise from photon shot noise, camera noise, and laser frequency uncertainty, all of which are discussed in this section.

The uncertainty $\Delta od^{(\ell)}_\mathrm{SN}$ for the $\ell$-th pixel due to photon shot noise is obtained by applying error propagation to Eq.~\ref{eq:od}  resulting in~\cite{Muessel2013}
\begin{align} \label{eq:ErrorOd}
	\notag \Delta od^{(\ell)}_\mathrm{SN} = \Biggr[ \left(\frac{\alpha^*}{\tilde{I}_{0}^{(\ell)}} + 1 \right)^2 & \left( \Delta \tilde{I}_{0}^{(\ell)} \right)^2 \\
 + &\left(\frac{\alpha^*}{\tilde{I}_\mathrm{p}^{(\ell)}} + 1 \right)^2 \left(\Delta \tilde{I}_\mathrm{p}^{(\ell)} \right)^2 \Biggr]^\frac{1}{2},
\end{align}
where $\Delta$ denotes the standard deviation of the quantities involved. While the noise contributions to $\Delta \tilde{I_j}$ in principle include photon shot noise, camera noise, and laser frequency uncertainty, the latter two are considered separately.

The shot noise is implemented from Eq.~\ref{eq:Intensity}, using the spread of the Poisson distribution for the photo-electrons $\Delta \mathcal{N}_\mathrm{el}^{(\ell)}=\sqrt{\mathcal{N}_\mathrm{el}^{(\ell)}}$ yielding~\cite{Genkina2015, Horikoshi2017, Lueckes_PhD}
\begin{equation}
\label{eq:DeltaIntensity}
\Delta \tilde{I_j} =
\frac{\sqrt{\mathcal{N}_\mathrm{el}^{(\ell)}}}{\mathcal{T}\eta} \frac{\hbar \omega}{I_\mathrm{s} \tau A}
= \sqrt{\tilde{I}_j} \sqrt{\frac{1}{\mathcal{T}\eta} \frac{\hbar \omega}{I_\mathrm{s} \tau A}}.
\end{equation}
An analysis of $\approx\!1000$ empty images (containing imaging light but no atoms) shows that shot noise is the predominant contributor to the noise of the individual pixel values, with an average spread of $\Delta od_\mathrm{SN}^{(\ell)} \approx 0.026$, or $\Delta N_\mathrm{SN}^{(\ell)} \approx 1.3$ per pixel, determined by the choice of imaging intensity and duration. Note that the shot noise for the individual pixel is completely uncorrelated to neighboring pixels. However, this analysis also shows indications of correlated noise attributed to camera effects. 

Camera noise typically includes photo response non-uniformity and read noise such as dark current effects, sense node reset noise, source follower noise, and ADC quantization noise~\cite{Konnik2014}. We did not specifically map out the pattern of the correlated camera noise but found that adding a global error $\Delta od_\mathrm{cam} = 2\times 10^{-4}$ (all pixels in the cloud region share the same value in each of the reconstructed images) to the shot noise reproduces the estimation error of the total atom number in the empty images.

Uncertainty in the laser frequency influences the transmitted light $\tilde{I}_\mathrm{p}$, which can be expressed using the Lambert W function~\cite{Pappa2011,Veyron2021arXiv}
\begin{align}\label{eq:Transmission}
    \tilde{I}_\mathrm{p}(\tilde{I}_0, od, \tilde{\delta}) &= \alpha^*(1 + 4\tilde{\delta}^2)\times \notag \\ 
    &W \Biggr[ \frac{\tilde{I}_0}{\alpha^*(1 + 4\tilde{\delta}^2)} \exp\left({\frac{\tilde{I}_0 -od}{\alpha^*(1 + 4\tilde{\delta}^2)}}\right) \Biggr].
\end{align}
The front factor $(1 + 4\tilde{\delta}^2)$ of the logarithmic term in Eq.~\ref{eq:od} can correct for this effect if $\tilde{\delta}$ is known. So, instead of evaluating the effect of variations in $\tilde{I}_\mathrm{p}$, one can equivalently explore the effect of variations in the front factor while assuming that $\tilde{I}_\mathrm{p}$ is independent of detuning. In our case, the frequency uncertainty arises from the precision of the laser lock based on Doppler-free atomic spectroscopy, measured to be $\Delta \delta \approx \SI{160}{\kilo \hertz}$ resulting in $\Delta \tilde{\delta}=0.027$. The lock typically operates on a time scale such that the frequency can be assumed to be constant within an imaging pulse of \SI{30}{\micro \second} duration.

The imaging frequency is resonant with the transition to maximize the $od$ signal, and thus, the first-order term in error propagation vanishes as $(\partial od/\partial \tilde{\delta})|_{\tilde{\delta}=0}=0$. However, some intuition can be gained from second-order error propagation~\cite{Crowder2020} revealing
\begin{align}
    \Delta od_\delta^{(\ell)} &=C'\frac{1}{2}\left(\frac{\partial^2 od^{(\ell)}}{\partial \tilde{\delta}^2}\right) (\Delta \tilde{\delta})^2\notag 
    \\ &=C' 4\alpha^*\ln\left(\frac{\tilde{I}^{(\ell)}_0}{\tilde{I}^{(\ell)}_\mathrm{p}}\right)(\Delta \tilde{\delta})^2.
\end{align}
Here, $C'= 1.35$ is a factor that aligns the result with the more accurate but computationally heavier Monte Carlo approach~\cite{Crowder2020}. The Monte Carlo approach, as opposed to the analytical error propagation given above, correctly captures the distortion of the Gaussian distribution of laser detuning within an expression exhibiting significant curvature (such as $od(\delta)$), and thus in the final implementation, we rely on the Monte Carlo approach. 

The expected estimation error of the total atom number is given by
\begin{align}\label{eq:DeltaN}
\Delta N = \frac{ A}{\sigma_0} \Biggr[ \sum_{\ell\in \mathrm{CR}} \left( \Delta od_\mathrm{SN}^{(\ell)} \right)^2  +&
 \left(n_\mathrm{px}\Delta od_\mathrm{cam}\right)^2 \notag\\
 +& \left(\sum_{\ell\in \mathrm{CR}} \Delta od_\delta^{(\ell)}\right)^2
 \Biggr]^{1/2} ,
\end{align}
where the sums are over all pixels $\ell$ in the \textit{cloud region} (CR) of the image and $n_\mathrm{px}$ is the number of pixels in this region. Since the contributions from camera noise and frequency uncertainty are completely correlated to all other pixels, the individual pixel contributions are added directly instead of in quadrature. Thus, effects that are negligible for the individual pixel can add up to a significant contribution when summing the signals from multiple pixels. 

Each of the three contributions in Eq.~\ref{eq:DeltaN} significantly affects the estimation error of the total atom number. For small cloud regions (i.e. when imaging cold clouds), these three effects contribute roughly equally. However, this is not the case for warmer clouds. In practice, the evaporation ramp leads to proportionality between $n_\mathrm{px}$ and $N$. Moreover, the contribution from imaging frequency uncertainty is roughly proportional to $N$, and as a result, both camera noise and imaging frequency uncertainty scale approximately linearly with $n_\mathrm{px}$, while shot noise scales as $\propto\!\sqrt{n_\mathrm{px}}$. Thus, when imaging warmer clouds and using larger cloud regions, camera noise, and imaging frequency uncertainty become the dominant effects, with the latter being the largest contributor. 

Equation~\ref{eq:DeltaN} offers insight into the factors contributing to the estimation error of $N$. However, in the final analysis, all these effects are implemented in the Monte Carlo simulation outlined in Sec.~\ref{Sec:EstimationErrorSim}, enabling the extraction of estimation errors for all cloud parameters: $N$, $\Nth$, $N_0$ and $T$.

% ------------ Finite number noise ------------------
\subsection{Finite number noise for the thermal cloud}
\label{sec:FiniteNumber}

This analysis assumes that the thermal cloud follows a Bose-enhanced Gaussian distribution for all temperatures. However, due to a finite number of thermal atoms, a perfect distribution never occurs, and Poisson noise is expected for the number of atoms per pixel. This finite number effect is easily simulated by randomly drawing the position of each atom based on the spatial probability function (Eq.~\ref{eq:Bose-enhanced}). Corrections due to bunching effects are discussed in Sec.~\ref{sec:additionalEffects}.

Apart from the extra noise on the individual pixels that impacts the thermal fitting, the finite number effect also causes binomial variations in the number of atoms present in the BEC region, and in the wings of the cloud (see Fig.~\ref{fig:cloudFitExample}), and even beyond the $3w_i$ wide cloud region. However, even if choosing the full cloud region to be $3\,w_i$ in radius does have a slight systematic effect on the total atom number, the statistical binomial variation in how many atoms are found inside and outside the region is negligible.

% ------------ Simulation of noise contributions------------------
\subsection{Simulation of noise contributions}
\label{Sec:EstimationErrorSim}

In the following, we describe a Monte Carlo simulation of optical density profiles of atom clouds, including the noise contributions discussed in Sec.~\ref{sec:odNoise} and \ref{sec:FiniteNumber}. This simulation allows for an analysis that mimics the experimental procedure described in Sec.~\ref{sec:cloudParams} and thus enables the extraction of estimation errors on $N_0$, $N_\mathrm{th}$, $N$, and $T$.

\begin{figure}
    \centering
    \includegraphics{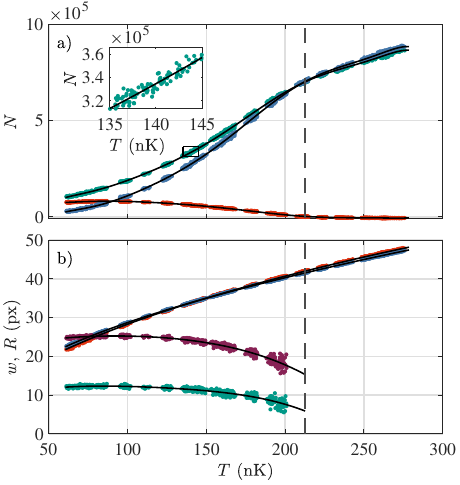}
    \caption{Polynomial fits of cloud parameters for all $\approx\!1000$ clouds in the analysis for the entire temperature range. a) Thermal atom number $N_\mathrm{th}$ (blue points), BEC number $N_0$ (red points), and total atom number  $N$ (green points) as a function of temperature $T$. The inset figure provides a closer view of the relationship between $N$ and $T$. b) The axial (red points) and radial (blue points) thermal cloud widths, and the axial (green points) and radial (purple points) Thomas-Fermi radii after time-of-flight expansion. The black lines indicated the polynomial fits used in the analysis. The vertical dashed lines indicate the critical temperature.}
    \label{fig:polyFits}
\end{figure}

The simulation was conducted for 40 temperatures covering the full experimental range. To simulate atom clouds at an arbitrary temperature, knowledge of the mean value of $N_0$, $N_\mathrm{th}$, the cloud widths, $w_i$, of the thermal cloud, and the Thomas-Fermi radii, $R_i$ as a function of temperature is required. These are obtained from higher-order polynomial fits $N_\mathrm{th}(T)$, $N_0(T)$, $w_i(T)$, and $R_i(T)$ to the experimental thermal atom number, the BEC number, and the widths, respectively, as a function of temperature for all $\approx\!1000$ clouds in the measurement. These fits are shown in Fig.~\ref{fig:polyFits}.

In the simulation, images corresponding to in Fig.~\ref{fig:cloudFitExample} are constructed as follows. For a given temperature, the thermal cloud widths are obtained from the polynomial fits and inserted in Eq.~\ref{eq:Bose-enhanced}, which is now interpreted as a probability function. Now, the position of each thermal atom is randomly drawn from the probability function, and they are placed on a grid corresponding to $400\times400$ pixels, which corresponds to more than $\pm4$ cloud widths of the warmest clouds. This produces simulated images with the appropriate spatial distribution and directly reproduces the finite number noise discussed in Sec.~\ref{sec:FiniteNumber}). The BEC is added according to a Thomas-Fermi distribution with the BEC atom number and radii determined from the polynomial fits.

Shot noise is randomly added to each pixel with a standard deviation of $\Delta od_{\mathrm{SN}} = 0.026$ obtained in Sec.~\ref{sec:odNoise}. To simplify the simulation, the magnitude is the same for all pixels regardless of the actual $od$. This can be justified since the optical density in the thermal cloud wings maximally differs from $od = 0$ to $od =1.5$, and thus, the expected difference in the magnitude of the shot noise for these two cases is on the level of $2\times 10^{-3}$. In the BEC region, larger $od$'s are recorded, leading to larger differences in the magnitude of shot noise. However, as will be clear later, the direct effect of shot noise on the BEC estimation is minimal. 

The completely correlated camera noise is added as a single random value from a Gaussian distribution to all pixels for each image with a standard deviation of $\Delta od_{\mathrm{cam}}  = 2.0\times10^{-4}$. Finally, the contribution from imaging frequency uncertainty is added by drawing a single frequency detuning, $\tilde{\delta}$, from a Gaussian distribution for each image with a standard deviation $\Delta \tilde{\delta} = 0.027$ and subtracting $4\alpha^*\tilde{\delta}^2\ln(\tilde{I}^{(\ell)}_0/\tilde{I}^{(\ell)}_\mathrm{p})$ from all pixels where $\ln(\tilde{I}^{(\ell)}_0/\tilde{I}^{(\ell)}_\mathrm{p})$ is determined individually for each pixel with $\tilde{I}^{(\ell)}_\mathrm{p}$ calculated from Eq.~\ref{eq:Transmission} with $\tilde{\delta}= 0$.

The simulated images of atom clouds are fitted with Eq.~\ref{eq:Bose-enhanced} using the least square method with the same error on all pixels. The settings for the \textit{cloud region} and the extraction of $N_0$, $N_\mathrm{th}$, $N$, and $T$ are identical to the evaluation of real clouds as described in Sec.~\ref{sec:cloudParams}, i.e. the \textit{wings} of the cloud are fitted to Eq.~\ref{eq:Bose-enhanced} to extract $N_\mathrm{th}$ and $T$. This thermal distribution is then subtracted from the \textit{BEC region}. $N_0$ is obtained by summing the remaining $od$ data in this region and $N$ by summing the raw data in the entire \textit{cloud region}.

There are three mechanisms contributing to the error of the BEC atom number. To gain further insight into these, we also record the summing error caused by $\Delta od^{(\ell)}$ in the BEC region, secondly, the variation of the thermal atom number in the BEC region (finite number effect), and, finally, the variation of the thermal fit evaluated in the BEC region.

The process of constructing images, adding noise, and fitting the thermal distribution was repeated 100,000 times for each temperature. This allows for the determination of the variances, covariances, and correlations between each of the measured quantities. In essence, the covariance matrix (see Sec.~\ref{sec:covarianceMatrix}) is obtained for each temperature.

\begin{figure}
    \centering
    \includegraphics[width=\columnwidth]{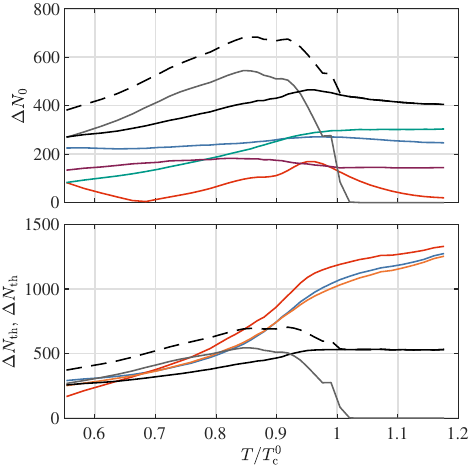}
    \caption{ Technical errors (standard deviation) of the BEC, thermal, and total atom numbers for various noise sources. The upper panel shows errors for the BEC atom number. (Blue) Error caused by uncertainty in the thermal fit in the BEC region. (Green) Error from the distribution of a finite number of thermal atoms in the BEC region. (Purple) Error from $\Delta od$ with the dominating contribution arising from laser frequency uncertainty. The lower panel shows the technical errors for the thermal and total atom numbers. (Blue) Estimation error of the thermal fit. (Orange) Estimation error of the total atom number. 
    In both panels, the red lines show the error inherited from the estimation error of the total atom number due to the method shown in Fig.~\ref{fig:residual}, while the grey lines show the \textit{system shift effect}. 
    The black solid lines show the total estimation error of the BEC and thermal atom number, respectively, and the dashed lines show the combination of all technical contributions, including the \textit{system shift effect}. 
    }
    \label{fig:EstimationErrors}
\end{figure}

\subsection{Cloud parameter estimation errors}

Figure~\ref{fig:EstimationErrors} shows the results of the simulation in Sec.~\ref{Sec:EstimationErrorSim}. The upper figure illustrates the estimation errors for the BEC atom number, where the primary contributions are the variation in the thermal fit evaluated in the BEC region (blue) and the binomial error in the BEC region (green) arising from the finite number of thermal atoms. The direct impact of errors in $od$ (purple) dominated by the effect of laser frequency uncertainty is slightly lower. The effect of the thermal fit and the laser frequency uncertainty remain relatively constant within the temperature range. The error due to finite number dominates at and above the critical temperature but decreases with temperature due to the reduction in thermal atoms. The remaining lines are discussed in subsequent sections. Note that the temperature axis is scaled with the critical temperature for an ideal gas, $T_\mathrm{c}^0$, which is not constant due to the loss of atoms during evaporative cooling.

The lower panel of Fig.~\ref{fig:EstimationErrors}, shows that the estimation error of the thermal atom number from fits (blue) is very similar to that of the total atom number (orange) obtained from summing the signal in the full cloud region. The decrease in the error towards lower temperature reflects the reduced number of atoms and the smaller cloud size with fewer pixels in the cloud region, hence leading to a smaller impact from $\Delta od$. Because of the decrease in the atom number, the relative error for both the total and the thermal atom number, in fact, increases. 

\begin{figure}
	\centering
	\includegraphics[width=0.96\columnwidth]{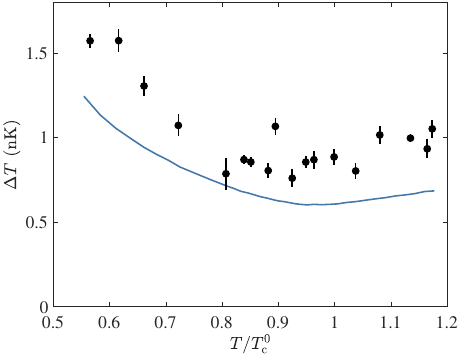}
	\caption{Spread of the temperature as a function of the reduced temperature $T/T_\mathrm{c}^0$. The black points represent the measured standard deviation of the temperature within a set of 60 clouds that were prepared with the same final rf. The error bars are extracted from Eq.~\ref{eq:varianceError} using error propagation. The blue line indicates the standard deviation obtained from the Monte Carlo simulation described in Sec.~\ref{Sec:EstimationErrorSim}. 
    }
	\label{fig:temperatureSpread}
\end{figure}

Figure~\ref{fig:temperatureSpread} displays the simulated estimation error for the temperature (blue line) which is slightly lower than the spread of the experimentally obtained temperature (black points). This difference may indicate the presence of preparation noise or further noise sources affecting temperature estimation. 

It is important to note that while the results mentioned above are specific to our imaging setup, the Monte Carlo simulation used to determine estimation errors can be applied to all analyses of density profiles obtained from absorption imaging.

\subsection{Additional noise sources}
\label{sec:additionalEffects}

The analysis presented above includes the most significant imaging effects that result in statistical errors in the parameter estimation. However, it should be mentioned that there may be other imaging effects that could potentially impact the statistical errors. Fortunately, most of these effects are primarily systematic and thus  have a low impact on statistical errors. This section provides a more detailed discussion of these additional imaging effects.

First, the imaging system has a finite resolution due to diffraction, which roughly corresponds to the length of one camera pixel. This is associated with a depth of field of $\approx\!\SI{32}{\micro \meter}$ while the full cloud size ($6 w_i$ diameter) varies from $0.5-1.1$~mm after time-of-flight, and hence, some of the atomic distribution must be out of focus. This primarily leads to systematic effects affecting temperature estimation. 

In addition, the atoms are accelerated during the imaging pulse due to momentum kicks from the probe beam followed by isotropic spontaneous emission. This introduces further defocussing with systematic consequences~\cite{Muessel2013}. The recoil kicks from spontaneously emitted photons also lead to a random walk of the atoms and diffusion of the cloud during imaging, resulting in further systematic effects~\cite{Muessel2013}. A simulation of our imaging setup shows that atoms typically move by less than one pixel and the resulting statistical effect is expected to be negligible.

Since $\approx\!400$ photons on average are absorbed per atom during imaging, the maximal Doppler shift at the end of the imaging pulse is $\approx\!\SI{3}{\mega \hertz}$. This will reduce the likelihood of absorption by only $20$\% for maximally accelerated atoms due to the high imaging power $(I_0 = 3.2~I_\mathrm{sat})$ and primarily cause systematic effects. These effects are partly compensated by the calibration factor, $\alpha^*$, in Eq.~\ref{eq:od}. The statistical component due to the stochastic absorption process is already implemented in the model of photon shot noise in Eq.~\ref{eq:DeltaIntensity}, and reflected in the larger relative errors in the intensity in the probe image, where some finite transmission through the cloud reduced the intensity reaching the camera.

To extract the cloud parameters the low-temperature limit of the Bose-enhanced Gaussian distribution Eq.~\ref{eq:Bose-enhanced} is used. This corresponds to setting the chemical potential to zero and, consequently, the fugacity to unity in the general Bose-enhanced distribution.
In the high-temperature limit, the general model describes a Gaussian distribution, but it becomes gradually more peaked when approaching the critical temperature. Equation~\ref{eq:Bose-enhanced} represents the low-temperature limit and the maximally peaked distribution, which is valid at the critical temperature. Using Eq.~\ref{eq:Bose-enhanced} for all temperatures introduces systematic effects since each measurement set contains a range of different total atom numbers and temperatures, and thus values for the fugacity. Since these two parameters are highly correlated, most of this effect is, however, caught by the \textit{correlation method}. Yet, if the correlation between temperature and total atom number is incomplete due to imperfect preparation (discussed in Sec.~\ref{sec:EnsembleShift}), the varying systematic effects can cause statistical effects. These effects, however, have shown to be insignificant in magnitude above the critical temperature.

Moreover, Eq.~\ref{eq:Bose-enhanced} is not the appropriate fitting function below the critical temperature as it does not account for the repulsive interaction between thermal and condensed atoms. Unfortunately, there is no applicable fitting model that takes thermal and condensed atoms into account and includes time-of-flight expansion. We have tested that an incorrect fitting model above the critical temperature had some systematic but negligible statistical effects.

The short time between the probe and the reference images, results in no observable fringes in the empty regions of the optical density profiles such as Fig.~\ref{fig:cloudFitExample}. However, when averaging all the residuals obtained by subtracting the thermal fit from the optical density profiles, within a measurement set, the mean residuals show some vertical fringes. The averaging is necessary to reduce the effect of shot noise. We ascribe these fringes to some degree of astigmatism in our imaging setup which is expected to have systematic consequences for the thermal fit, but due to their static nature, the statistical effect is expected to be low. 

When analyzing a single optical density profile and subtracting both the thermal fit and the mean residual mentioned above, it should be possible to observe the pure shot noise and distribution noise. However, these residuals contain more structure than expected from the simulation in Sec.~\ref{Sec:EstimationErrorSim}, even when considering some finite resolution of the imaging system. This structure is a stochastic pattern and differs for each realization. It is thus plausible that these structured deviations result in additional errors in the thermal fit compared to the simulation.

We speculate this structure could be caused by lensing effects in the atomic cloud. Another possible explanation is the effect of bunching. In our Monte Carlo method, we did not consider statistical correlations between the atoms, i.e. in the simulation, the position of each atom is drawn independently of previously drawn positions. This approach does not account for the bunching effect in thermal atoms~\cite{Burt1997}. However, the averaging of each column density is expected to reduce the potential role of bunching in the analysis. Both of these effects are expected to be small but may be important contributors to the structured residuals.

% ###############################################################################
% ------------------ Fluctuation measurement noise contributions ----------------
% ###############################################################################

\section{Fluctuation measurement noise contributions}
\label{sec:FlucsMeasurementNoise}

The analysis of atom number fluctuations in BECs employs techniques beyond the regular determination of atom numbers. In particular, the \textit{correlation method} to eliminate the effect of slow drifts can add a technical variance. Moreover, remaining uncertainties in the atom number preparation also critically impact the measurement of atom number fluctuations. This section discusses these two technical noise sources, which are specific for the measurement of atom number fluctuations. 

% ------------------ Error inherited due to correlation method ----------------

\subsection{Error inherited due to \textit{correlation method}}\label{sec:ErrorFromN}

In the analysis of atom number fluctuations, the technical drift within a measurement set is eliminated by subtracting a linear fit from the relevant atom number ($N_0(N)$ or $N_\mathrm{th}(N)$) as discussed in Sec.~\ref{sec:fluctmeasure} and shown in \figref{fig:residual}. This \textit{correlation method} was essential for the determination of BEC atom number fluctuations~\cite{Kristensen2019,Christensen2021}.

However, due to the subtraction, the residuals of $N_0$ and $N_\mathrm{th}$ inherit any estimation error of the total atom number $N$. Such an estimation error, $\delta N$, causes a shift of $N_k$ equal to $\delta N_k = - \frac{\mathrm{d}N_k}{\mathrm{d}N}\delta N$ where $N_k$ can be $N_0$ or $N_\mathrm{th}$. The resulting technical errors are shown in Fig.~\ref{fig:EstimationErrors} as red lines for both the BEC (upper panel) and the thermal cloud (lower panel). The nontrivial behavior of these curves is caused by the macroscopic correlation $N_0(N)$ and $N_\mathrm{th}(N)$.

The slope $\mathrm{d}N_0/\mathrm{d}N$ is small in magnitude for all temperatures because it generally requires the removal of several thermal atoms to gain one atom in the BEC. Therefore, this effect is not very significant for the BEC. On the other hand, $\mathrm{d}N_\mathrm{th}/\mathrm{d}N=1-\mathrm{d}N_0/\mathrm{d}N$ is always close to one, and the effect is very important for the thermal clouds.

% ------------------ system shift effect ---------------------------------

\subsection{System shift effect}
\label{sec:EnsembleShift}

The second effect specific to measuring atom number fluctuations is due to the imperfect preparation of the atom clouds. In the semi-ideal model~\cite{Naraschewski1998}, the BEC atom number is given by
\begin{align} \label{eq:semi-ideal}
\frac{N_0}{N} = 1- \left(\frac{T}{T_\mathrm{c}}\right)^3 - 
\eta \frac{\zeta(2)}{\zeta(3)}\left(\frac{T}{T_\mathrm{c}^0}\right)^2 
\left(\frac{N_0}{N}\right)^{2/5}.
\end{align}
Here, the scaling parameter is defined as
\begin{equation}
    \eta = \frac{1}{2} \zeta(3)^{1/3}\left( 15 N^{1/6} a  \right)^{2/5}
\end{equation}
where the scattering length $a$ is scaled in units of $\sqrt{\hbar/m\overline{\omega}}$. Equation~\ref{eq:semi-ideal}, shows the dependence of the BEC atom number on the total atom number, the temperature, and the trap geometry. 

Experimentally, the total atom number and the temperature are highly correlated due to the evaporation process. This effect was discussed in Sec.~\ref{sec:ExperimentalSetup} and shown in Fig.~\ref{fig:residual}, which dictates the subtraction of a linear fit before evaluating the variance. However, the relationship between $T$ and $N$ is not completely deterministic as shown in the inset of Fig.~\ref{fig:polyFits}. According to Eq.~\ref{eq:semi-ideal}, any preparation noise in the total atom number, which is uncorrelated to the temperature, can cause variance in the BEC atom number, which is not captured by the \emph{correlation method}. This effect is called the \emph{system shift effect}, as it corresponds to probing systems with different initial conditions. It is straightforward to quantify this effect by solving Eq.~\ref{eq:semi-ideal} numerically for combinations of $T$ and $N$, where uncorrelated atom number variations are included.

The correlation coefficient between $T$ and $N$ within measurement sets at a given final radio frequency provides intuition of their experimental relation. Indeed, this coefficient is not $1$, and further, it decreases towards colder sets, indicating that sets exposed to more evaporative cooling after the stabilization step have a relatively larger uncorrelated atom number variance. 

To extract the uncorrelated atom number variance necessary for estimating the \emph{system shift effect}, we divide the variance in the total atom number into two parts, one that is completely correlated and one completely uncorrelated to the temperature. Assuming that no further effects influence the correlation between $N$ and $T$, it is possible to extract the uncorrelated spread of the total atom number for a measurement set
\begin{equation}
    \Delta N_\mathrm{uncorr} = \sqrt{\Delta N^2 - \Delta N_\mathrm{corr}^2}
\end{equation}
where $\Delta N^2$ is the experimentally observed variance of the total atom number and $\Delta N_\mathrm{corr} = \mathrm{cov}(N,T)/\Delta T$ is the correlated atom number spread, with $\mathrm{cov}(N,T)$ the observed covariance between $N$ and $T$ and $\Delta T$ the observed temperature spread. For each measurement set at a given final rf, one can thus obtain $\Delta N_\mathrm{uncorr}$ and $\Delta N_\mathrm{corr}$. Since estimation errors for $N$ and $T$ are not considered this variance sets an upper limit to the uncorrelated variance in the total atom number. Therefore, we use the procedure described above to find the temperature dependence of $\Delta N_\mathrm{uncorr}$, but we introduce a global \textit{preparation noise parameter}, $\epsilon$, which scales this variance.

In brief, the simulation is carried out as follows. For each temperature $T$, the average number $N$, and the variances $\Delta N_\mathrm{uncorr}$ and $\Delta N_\mathrm{corr}$ are obtained from global fits to the data. The variance $\Delta N_{uncorr}$ is scaled with the preparation noise parameter, $\Delta N_{uncorr} = \epsilon \times N_{\mathrm{uncorr}}$ and $\Delta N_\mathrm{corr}$ is chosen correspondingly to fix $\Delta N$. Then 100,000 random atom numbers, $N_j$ are drawn from the Gaussian distribution $\mathcal{G}(\mu = N,\: \sigma =  \Delta N_{\mathrm{corr}})$. For all $N_j$, the corresponding temperature is found $T_j = T(N_j)$. In addition, 100,000 uncorrelated atom spreads are now drawn from  $\mathcal{G}(\mu = 0,\: \sigma = \Delta N_{\mathrm{uncorr},i})$ and added element-wise to a final $N_j$.

The semi-ideal model is then solved for each realization of $T_j$ and $N_j$. All resulting $N_{0,j}$ are plotted against $N_j$, and the \textit{correlation method} in Fig.~\ref{fig:residual} is carried out to evaluate the variance. This is repeated for all temperatures $T$ and for all preparation noise parameters $0\leq \epsilon \leq0.26$ in steps of $0.01$. The result of this simulation for $\epsilon=0.18$ is shown in Fig.~\ref{fig:EstimationErrors} as grey lines for the BEC (upper panel) and the thermal cloud (lower panel)~\footnote{Note, that the kink in the curve just below the critical temperature can be attributed to the fit of $N_0$ as a function of $N$ close to the critical temperature and is not of fundamental nature.}. Note, that the effect on the BEC and thermal atom number is completely anticorrelated by construction, and therefore, the resulting spread is exactly the same. The consequences of the combined noise analysis are discussed in the following section.

% ###############################################################################
% ------------------ Complete fluctuation and noise analysis ----------------
% ###############################################################################

\section{Complete fluctuation and noise analysis}
This section combines the noise contributions discussed in Sec.~\ref{Sec:CloudAnalysis} and Sec.~\ref{sec:FlucsMeasurementNoise} and analyses their effect on the size of the fundamental atom number fluctuations. First, we present the formalism for the combination of errors and apply it to our analysis.

% ----------- Combination of errors and the covariance matrix --------------

\subsection{Covariance matrix formalism}
\label{sec:covarianceMatrix}

Since the noise contributions are not independent of each other, it is necessary to use the covariance matrix to combine them, rather than simply adding the variances. For a set of  variables, $x_i$ (such as all the technical noise sources), information about the variances and covariances are given in the covariance matrix, $\mathbf{\Sigma}_x$, defined as $\mathrm{cov}(x_i, x_j) = \langle(x_i-\mu_i)(x_j-\mu_j)\rangle$ or for a discrete sample $\mathrm{cov}(x_i, x_j) = 1/(n-1) \sum_{k=1}^n \left( x_i^{(k)}-\overline{x_i}\right) \left(x_j^{(k)}-\overline{x_j} \right)$~\cite{Barlow1997}. If a set of variables, $f_l$ (such as the raw measurement of the BEC and thermal atom number), depends on $x_i$, the covariance matrix, $\mathbf{\Sigma}_f$, can be calculated as 
\begin{equation}\label{eq:ErrorPropagation}
    \mathbf{\Sigma}_f = \mathbf{J\Sigma}_x\mathbf{J}^\mathrm{T}
\end{equation}
with 
\begin{equation}
    J_{li} =  \left( \frac{\partial f_l}{\partial x_i} \right).
\end{equation}
Setting up the noise sources in this well-known formalism is a convenient tool for adjusting individual contributions if the measured or simulated noise contributions do not fully reflect the total observed noise.

\subsection{Combination of noise contributions}

To combine the estimation errors (Sec.~\ref{Sec:EstimationErrorSim}), the inherited errors from the measured total atom number $N$ (Sec.~\ref{sec:ErrorFromN}), and the \textit{system shift effect} (Sec.~\ref{sec:EnsembleShift}), one defines the total technical noise variables as the sum of all noise contributions according to
\begin{align}
    N_{0,\mathrm{tech}} &= N_{\mathrm{0,est}} - \frac{\mathrm{d}N_0}{\mathrm{d}N} N_{\mathrm{est}} + N_{\mathrm{ss}}(\epsilon)\\
    N_\mathrm{th,tech} &= N_\mathrm{th,est} - \frac{\mathrm{d}N_\mathrm{th}}{\mathrm{d}N} N_{\mathrm{est}} - N_{\mathrm{ss}}(\epsilon)
\end{align}
where $N_{\mathrm{0,est}}$, $N_{\mathrm{th,est}}$, and $N_\mathrm{est}$ are the estimation errors of the BEC, thermal, and total atom number, respectively, and $N_\mathrm{ss}(\epsilon)$ is the \textit{system shift effect}. Importantly, $N_\mathrm{ss}(\epsilon)$ depends on the choice of $\epsilon$. Note that these are all stochastic variables with a mean value equal to zero. From this, one can define
\begin{equation}
    \mathbf{J} =
    \begin{pmatrix}
    1 & 0 & -\frac{\mathrm{d}N_0}{\mathrm{d}N} & 1 \\[0.3cm]
    0 & 1 & -\frac{\mathrm{d}N_\mathrm{th}}{\mathrm{d}N} & -1
    \end{pmatrix}
\end{equation}
and set up the covariance from our simulation of estimation errors 
\begin{equation}\label{eq:covarianceMatrix}
    \mathbf{\Sigma} =
    \begin{pmatrix}
    \Delta N_\mathrm{0,est}^2 & c_\mathrm{0,th} & c_\mathrm{0,tot} & 0 \\[0.3cm]
    c_\mathrm{0,th}  & \Delta N_\mathrm{th,est}^2 & c_\mathrm{th,tot} & 0\\[0.3cm]
    c_\mathrm{0,tot}  & c_\mathrm{th,tot} & \Delta N_\mathrm{sum}^2 & 0\\[0.3cm]
    0 & 0 & 0 & \Delta N_\mathrm{ss}^2(\epsilon)
    \end{pmatrix}.
\end{equation}
The covariances $c_{i,j}$ between the estimations errors extracted from the simulation in Sec.~\ref{Sec:EstimationErrorSim}. Here, $i,j$ stands for the BEC, the thermal and total atom number using  $0$, $th$, and $tot$, respectively. Finally, the total technical noise contributions and their covariances can be calculated.
\begin{equation}\label{eq:technicalNoise}
    \mathbf{J\Sigma J}^\mathrm{T} =
    \begin{pmatrix}
    \Delta N_\mathrm{0,tech}^2 & c_\mathrm{0,th, tech} \\[0.4cm]
    c_\mathrm{0,th, tech} & \Delta N_\mathrm{th,tech}^2
    \end{pmatrix}
\end{equation}
This allows us to plot the total technical error $\Delta N_\mathrm{0,tech}$ and  $\Delta N_\mathrm{th,tech}$ in Fig.~\ref{fig:EstimationErrors}. For the preparation noise parameter $\epsilon=0$, the system shift disappears, and the total technical error includes only estimation errors (solid black lines). The dashed black lines represent the total technical error when choosing the best estimate for $\epsilon$ (see Sec.~\ref{sec:FittingFlucs}). Note importantly that the strong correlation between $N_\mathrm{est}$ and $N_\mathrm{th,est}$, as well as the minus sign in front of  $\mathrm{d}N_\mathrm{th}/\mathrm{d}N$ in $\mathbf{J}$, lead to a partial cancellation of estimation error on $N_\mathrm{th}$ from fitting, and the error inherited from $N$. This results in the total error on $N_\mathrm{th}$ being smaller than the sum of the individual errors, highlighting the importance of using the covariance matrix in the noise analysis.

\begin{figure}
    \centering
    \includegraphics[width=\columnwidth]{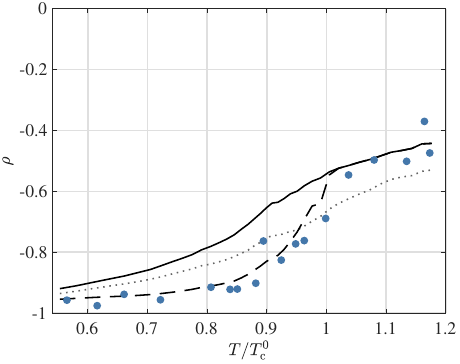}
	\caption{Correlation between the BEC and the thermal atom number. Blue points indicate the measured correlation. The dotted line is the expected correlation from estimation errors only without including the \textit{system shift effect} (see Sec.~\ref{sec:EnsembleShift}) extracted directly from the Monte Carlo simulation. The solid black line is the corrected correlation from estimation errors only as discussed in Sec.~\ref{sec:FittingFlucs}. The dashed black line represents the corrected noise model, including the estimated impact of the \textit{system shift effect}.}
	\label{fig:correlations}
\end{figure}

Moreover, the covariance term can be converted to a correlation, $\rho = c_\mathrm{0,th, tech}/(\Delta N_\mathrm{0,tech} \Delta N_\mathrm{th,tech})$. These correlations are shown in comparison with experimental results in Fig.~\ref{fig:correlations}. First, the preparation noise parameter, $\epsilon$, of the \textit{system shift effect} is set to zero and plotted as a dotted black line. The experimental results are the measured correlation between the residuals of the BEC atom numbers and the thermal atom numbers shown in Fig.~\ref{fig:residual}(b) and (d) for individual sets of 60 clouds that were prepared under the same conditions. Note that it is not the correlation between the raw measurement of the BEC and thermal cloud, which is caused by technical drift of the evaporation. The residual, after the \textit{correlation method}, should be cleaned for this.

Importantly, the noise analysis shows a strong anticorrelation in agreement with the experimental data. This indicates that the observed anticorrelation is primarily introduced by the analysis technique rather than the inherent fluctuations of the atom number, as will be discussed in Sec.~\ref{sec:conclusion}.

Since the \textit{system shift effect} is absent and the fundamental fluctuations have not yet been considered, there is an expected disagreement between the measured correlations and the simulated result below the critical temperature. However, above the critical temperature, any differences indicate deficiencies in the assumptions for the Monte Carlo simulation of estimation errors. This is treated in the next section.

% ------------------ Results ------------------------------
\subsection{Noise analysis in fluctuation measurements}
\label{sec:FittingFlucs}

The noise analysis outlined in the previous sections and their combination using the covariance matrix lays the basis for a full analysis of the noise sources affecting the measurement of atom number fluctuations in quantum degenerate Bose gases. However, in practice, two problems arise that complicate this analysis. On the one hand, the noise analysis does not fully capture the noise observed in the absence of a BEC. Secondly, the size of the technical noise parameter, $\epsilon$, which affects the impact of the \textit{system shift effect}, needs to be determined.

\begin{figure}
	\centering
		\includegraphics[width=0.95\columnwidth]{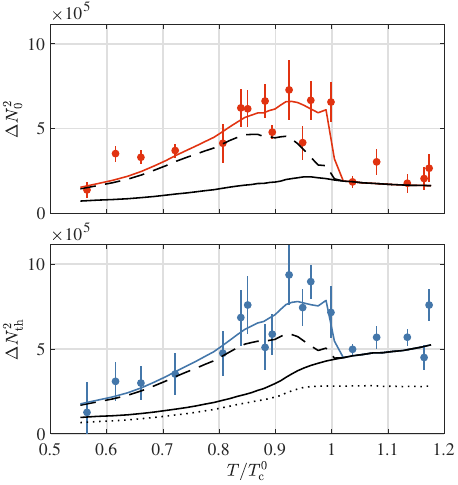}
	\caption{Variance of the BEC (upper panel) and the thermal atom number (lower panel) as a function of temperature. Data points represent the measured variance,  where the error bars are obtained from Eq.~\ref{eq:varianceError}. The solid black lines indicate the variance from estimation errors. In the thermal case, a dotted line indicates the estimation error extracted from the Monte Carlo simulation, while the solid line represents the correction of the simulation discussed in Sec.~\ref{sec:FittingFlucs}. Additionally, the dashed black lines represent the total technical noise, including the estimated impact of the \textit{system shift effect}.
 }
	\label{fig:fluctuations}
\end{figure}

\subsubsection{Thermal cloud noise analysis}
\label{sec:thermalnoise}

When examining the variance of the atom number in the thermal cloud above the critical temperature in Fig.~\ref{fig:fluctuations}, the simulated estimation error of the thermal atom number lies below the experimental one. Moreover, the simulated anticorrelation between the BEC and the thermal atom number is too high compared to the measured quantities as shown in Fig.~\ref{fig:correlations}. Note that there is no BEC at $T/T_c^0>1$ such that only estimation errors are involved.

Three factors impact the thermal atom number variance above the critical temperature; the estimation error on the thermal fit, the error in total atom number, and the covariance between these two. The total atom number variance should be captured well by our simulation result. This is supported by the observation that an increase in the shot noise, camera noise, or laser uncertainty contributions to $\Delta od$, does not significantly change the above-mentioned differences. However, we strongly suspect that the structured residuals discussed in Sec.~\ref{sec:additionalEffects} might cause larger thermal fitting errors and less correlation to the total atom number.

The above reasoning motivates an adjustment of the noise model at the level of the covariance matrix.  We find that this can be achieved by increasing the estimation error of the thermal atom number by \SI{20}{\percent} and by reducing the anticorrelation of the thermal to the total atom number by \SI{4}{\percent} . This adjustment showcases the power of using the covariance matrix as a tool in this analysis. Importantly, this only affects the variance of the thermal number of atoms and now provides a realistic background value. The solid lines in Fig.~\ref{fig:fluctuations} (lower panel) and Fig.~\ref{fig:correlations} represent these changes.

\subsubsection{Technical noise parameter determination}
\label{sec:technoiseparam}

The second issue to be addressed is the value of the  technical noise parameter, $\epsilon$, which sets the size of the \textit{system shift effect}. To determine the best estimate of $\epsilon$ we use the following method.  

For each value of $\epsilon$ the corresponding variance from the \textit{system shift effect} is inserted in $\mathbf{\Sigma}$ (Eq.~\ref{eq:covarianceMatrix}), and the total technical variance is calculated using Eq.~\ref{eq:technicalNoise} for each temperature, corresponding to dashed lines in Fig.~\ref{fig:fluctuations}. This variance is subtracted from the measured variance of the BEC and thermal atom number and the remaining signal is fitted with a model adapted from~\cite{Christensen2021}
\begin{equation}\label{eq:fitModel}
	\Delta N_{j,\mathrm{model}}^2(\Delta N_{j,p}^2,T_\mathrm{p}, T) = \left(f\ast g\right)(T)
\end{equation}
where $f$ and $g$ are given by
\begin{align}
	f(T) &= \Delta N_{j,\mathrm{p}}^2 \left(\frac{T}{T_\mathrm{p}}\right)^3 \Theta(T_\mathrm{p}-T), \label{eq:fitModelUnconvo}\\
	g(T) &= \mathcal{N}(T,\sigma_T). \label{eq:tempSpread}
\end{align}
The two free parameters are the peak atom number variances $\Delta N_{0,\mathrm{p}}^2$, $\Delta N_\mathrm{th,p}^2$. The temperature of peak fluctuations $T_\mathrm{p}$ is set at the value already determined in~\cite{Christensen2021}. Here, $\mathcal{N}(T,\sigma_T)$ is a normal distribution centered on the temperature $T$ with a standard deviation $\sigma_T$ given by the median of the measured temperature variation. The convolution of Eq.~\ref{eq:fitModelUnconvo} with Eq.~\ref{eq:tempSpread} accounts for the temperature spread within a set of 60 clouds.

The fit to the remaining signal is based on the sum of the normed square residuals, $\chi^2 = \sum_{i=1}^n [(\Delta N_j^i)^2(\epsilon) - \Delta N_{j,\mathrm{model}}^2(\Delta N_{j,p}^2,T_\mathrm{p},T_i) ]^2/e_i^2$, for the entire model with $\Delta N_{j,p}^2$ as the free parameter.  Here, $(\Delta N_j^i)^2(\epsilon)$ represents the measured variances after subtracting technical noise, for the BEC $(\Delta N_0^i)^2(\epsilon)$ or the thermal cloud $(\Delta N_\mathrm{th}^i)^2(\epsilon)$ with the errors $e_i$ on the measured variances, and $\Delta N_{j,\mathrm{model}}^2(\Delta N_{j,p}^2,T_i)$ is the prediction of fundamental fluctuations at the same temperatures. We determine the optimal value of $\epsilon$ by finding the value which minimizes $\chi^2$.

\begin{figure}
    \centering
    \includegraphics{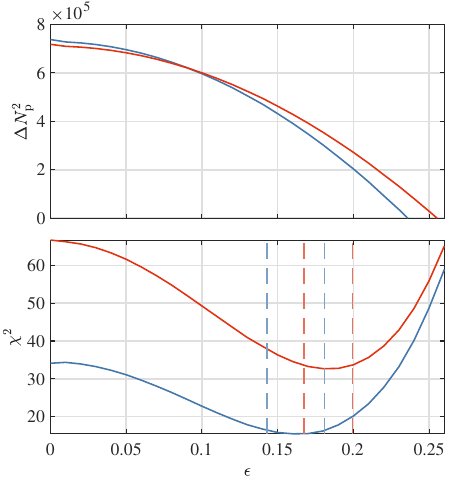}
    \caption{The upper figure shows the extracted fundamental peak fluctuations for different choices of the preparation noise parameter, $\epsilon$, which controls the impact of the \textit{system shift effect} described in Sec.~\ref{sec:EnsembleShift}. The lower figure shows the $\chi^2$ for the noise model as a whole, which includes the fit of the fundamental fluctuations. Red lines represent the analysis of the BEC data, while the thermal case is shown in blue lines.  In the lower figure, the vertical dashed lines mark the 68\% confidence interval for the most probable $\epsilon$ for both the BEC and thermal cloud.}
    \label{fig:MaxFlucsVsFraction}
\end{figure}

Figure~\ref{fig:fluctuations} includes the sum of the fit and all technical noise contributions with the optimal choice of $\epsilon$ for the BEC and the thermal fluctuations. This choice of $\epsilon$ is detailed in Fig.~\ref{fig:MaxFlucsVsFraction} which shows the peak fluctuations (upper panel), and $\chi^2$ (lower panel), as a function of the preparation noise parameter, $\epsilon$, for the BEC and the thermal cloud. 

Importantly, Fig.~\ref{fig:fluctuations} shows very good agreement between observed data and the overall noise model combined with a fit to the fundamental fluctuations. For the chosen model of the \textit{system shift effect}, fundamental fluctuations are necessary to explain the observed atom number variances. However, it is clear that extracted fundamental fluctuations are sensitive to the level of the \textit{system shift effect}. The peak fluctuation amplitude at the best estimate for $\epsilon$ is $\Delta N_\mathrm{p,0}^2 =(3.7\pm7)\times10^5$  and $\Delta N_\mathrm{p,th}^2 =(3.2\pm8)\times10^5$. Compared to earlier analysis~\cite{Christensen2021}, the estimate based on $N_0$ is lower by \SI{41}{\percent}. 

Figure~\ref{fig:MaxFlucsVsFraction} shows that the best estimates for the preparation noise parameters are $\epsilon_0 = \SI{0.184\pm0.016}{}$ and $\epsilon_\mathrm{th} = \SI{0.163\pm0.019}{}$ extracted from the BEC and thermal atom number variance, respectively, with the weighted mean value for $\overline{\epsilon}=\SI{0.175\pm0.012}{}$. The error corresponds to the distance between the $\chi_\mathrm{min}^2$ and the $\chi_\mathrm{min}^2+1$ values. Within this error, there is good agreement between the two estimates for the preparation noise parameter. 

Overall, it confirms that the fundamental fluctuations can indeed be measured despite the presence of the \textit{system shift effect}. However,  Fig.~\ref{fig:MaxFlucsVsFraction} also demonstrates that sufficiently high values of the preparation noise parameters will impede the observation of fundamental peak fluctuations. This means that measuring atom number fluctuations requires an experiment with low technical noise rather than relying on post-selection of data.

% ------------------ Conclusion ------------------------------
\section{Conclusion}
\label{sec:conclusion}

Motivated by the recent observation of atom number fluctuations in weakly interacting Bose gases, this paper presents a comprehensive statistical analysis of uncertainties in the preparation and parameter estimation of partially condensed Bose gases. In particular, the combined analysis of the condensed and thermal components shows the presence of fluctuations in both components. The analysis allows for improved estimation of the peak BEC atom number fluctuations and provides a guide for future measurements of atom number fluctuations.

Our statistical analysis consists of two parts. First, the estimation errors generally relevant for experiments using absorption imaging are analyzed. In the second part, effects specifically relevant to the measurement of atom number fluctuations are treated. These contributions are combined using the covariance matrix approach, which importantly allows for the phenomenological adjustment of individual contributions. Finally, the impact of this noise analysis on the measurement of atom number fluctuations is discussed.

The first part of the analysis shows that a Monte~Carlo method can be used to analyze the effect of photon shot noise, camera noise, laser frequency uncertainty, finite number effects, and the image evaluation procedure. Thus, a detailed understanding of the estimation errors of the BEC and thermal cloud atom number can be gained.

The theoretically obtained estimation errors for the BEC atom number match the background value (for $T/T_c^0>1$) of the observed BEC atom number variance $\Delta N_0^2$ extremely well. However, our simulation does not capture the estimation errors of the larger thermal cloud equally well. Thus, the background value (for $T/T_c^0>1$) of the thermal atom number variance $\Delta N_\mathrm{th}^2$ is underestimated. This indicates that further noise sources will have to be included in future analysis. In particular, we suspect that an observed structured stochastic pattern in the imaging noise could significantly contribute. 

Importantly, our analysis shows that the different contributions to estimation errors are highly correlated for both the BEC and the thermal atom numbers. Thus, they can not simply be added in quadrature and require an analysis in terms of the covariance matrix, which also enables a phenomenological treatment of the background thermal atom number variance. 

Moreover, our analysis allows for an evaluation of the error in the temperature estimation. The simulated spread of the temperature is in close agreement with the experimental result.

The second part of our analysis considers the effects of the \textit{correlation method} and takes imprecision in the preparation of the atomic ensembles into account. In particular, the \textit{correlation method} is shown to lead to an additional error inherited from the estimation error of the total atom number. This only leads to a minor effect on the BEC atom number variance but contributes considerably to the variance of the thermal atom number. The method also introduces an inherent anticorrelation between the BEC and thermal atom number. 

Finally, our analysis shows how noise in the preparation of the atomic samples affects variations of the BEC fraction - here called the \textit{system shift effect}. In combination with the \textit{correlation method}, this gives rise to additional correlated noise, which emerges just below the critical temperature, increases towards lower temperature peaking at $T/T_c^0\approx\!0.85$, and falls off slowly. Due to its slow onset as temperature decreases, it does not preclude the observation of fundamental atom number fluctuations but affects the extraction of their amplitude. However, in scenarios with significantly higher preparation noise than in our experiment, the \emph{system shift effect} becomes predominant, rendering it impossible to measure the fundamental fluctuations through data post-selection in experiments with large technical atom number fluctuation but precise detection.

The complete noise analysis also shows that the parameter extraction method leads to the strong observed anticorrelation between the thermal and the BEC atom numbers. Thus, the measurement of atom number fluctuations in the thermal cloud is not an independent observation of fluctuations in a quantum degenerate Bose gas. Instead, it should be viewed as a complementary method to obtain information on the fluctuations, albeit at a larger background variance. This also shows that the expected anticorrelation between the thermal and the BEC atom numbers due to fundamental fluctuations can not be observed directly in our experiments. 

When considering all technical noise sources, the estimated peak atom number fluctuations in the BEC are reduced by \SI{41}{\percent} compared to earlier analyses that relied on simpler assumptions for technical noise. This new estimate sets a benchmark for theoretical work, however, a direct comparison between experimental measurements and theoretical predictions is currently not possible due to the absence of theoretical results for typical experimental conditions. Such a comparison is further complicated by the fact that theory deals with ideal, mathematically defined ensembles, which do not necessarily represent an experimental realization.

In summary, a full noise analysis of the preparation and parameter extraction techniques for the investigation of atom number fluctuations in quantum degenerate gases was performed. This analysis shows the non-trivial, correlated nature of the noise contributions in the BEC and thermal atom numbers. In particular, it shows that a complementary observation of the fluctuations is possible in the thermal atom number and that the peak BEC atom number fluctuations are reduced due to an improved model of the background noise. 

\section{Outlook}
\label{sec:outlook}

In future experimental work, the main focus will be the reduction of preparation noise, which could be achieved by placing the atom number stabilization later in the experimental sequence. Moreover, an increased repetition rate of the sequence offers a direct approach to reduce the effect of drifts. The observed technical correlations between the BEC and thermal atom numbers could be reduced by using Bragg pulses to separate the two components and thus obtain independent measurements. These improvements will drastically enhance the measurement of atom number fluctuations and potentially allow for the observation of particle exchange between the two components and their expected atom number anticorrelation.

Finally, the results presented here provide a quantitative benchmark for further theoretical developments. In particular, a better method to simulate larger, interacting systems and a relaxation of the mathematically defined statistical ensembles are needed. This may provide a more detailed model of the fluctuations as a function of temperature under typical conditions, which will be highly beneficial for future experiments. The recently proposed Fock State Sampling method~\cite{Kruk2023} may provide a first step to achieve these goals.

\section{Acknowledgments}
\label{sec:acknowledgments}

Thanks to Maximillian Baron for valuable discussions.
We acknowledge support from the Danish National Research Foundation through the Center of Excellence “CCQ” (DNRF152), by the Novo Nordisk Foundation NERD grant (Grantno. NNF22OC0075986), and by the Independent Research Fund Denmark (Grantno. 0135-00205B).
K. P. acknowledges support from the (Polish) National Science Center Grant No. 2019/34/E/ST2/00289.
K. R. acknowledges support from the (Polish) National Science Center
Grant No. 2021/43/B/ST2/01426. 
Center for Theoretical Physics of the Polish Academy of
Sciences is a member of the National Laboratory of Atomic, Molecular and Optical Physics
(KL FAMO).

\bibliography{thermalFluctuations}

\end{document}